\UseRawInputEncoding
\documentclass{optica-article}

\journal{opticajournal} % for journals or Optica Open

\articletype{Research Article}

\usepackage{lineno}

\begin{document}

\title{Silicon Nitride Photonic Waveguide-Based Young's Interferometer for Molecular Sensing}

\author{Sahar Delfan,\authormark{1,2} Mohit Khurana,\authormark{1,2,*} Zhenhuan Yi,\authormark{1,2} Alexei Sokolov,\authormark{1,2} Aleksei
M. Zheltikov,\authormark{1,2} and Marlan O. Scully\authormark{1,2,3,4}}

\address{\authormark{1} Department of Physics and Astronomy, Texas A\&M University, College Station, TX 77843, USA\\
\authormark{2}Institute of Quantum Science and Engineering, Texas A\&M University, College Station, TX  77843, USA\\
\authormark{3}Baylor University, Waco, Texas 76704, USA\\
\authormark{4}Princeton University, Princeton, New Jersey 08544, USA}

\email{\authormark{*}mohitkhurana@tamu.edu} %% email address is required; see note below about the corresponding author designation

% use {asbstract*} to suppress the copyright line. Copyright information will be added in production

\begin{abstract*} Devices based on photonic integrated circuits play a crucial role in the development of low-cost, high-performance, industry-scale manufacturable sensors. We report the design, fabrication, and application of a silicon nitride waveguide-based integrated photonic sensor in Young's interferometer configuration combined with Complementary Metal-Oxide-Semiconductor (CMOS) imaging detection. We use a finite-difference time-domain method to analyze the performance of the sensor device and optimize the sensitivity of the fundamental transverse-electric (TE) mode. We develop a low-cost fabrication method for the photonic sensor chip, using photolithography-compatible dimensions, and produce the sensing region with wet-etching of silicon dioxide. We demonstrate the sensor's functioning by measuring the optical phase shift with glucose concentration in an aqueous solution. We obtain consistent interference patterns with fringe visibility exceeding 0.75 and measure the phase differences for glucose concentrations in the $10 \,\, \mu g/ml$ order, corresponding to the order of $10^{7}$ molecules in the sensing volume. We envision extending this work to functionalized surface sensors based on molecular binding. Our work will impact biosensing applications and, more generally, the fabrication of interferometric-based photonic devices.
\end{abstract*}

%%%%%%%%%%%%%%%%%%%%%%%%%%  body  %%%%%%%%%%%%%%%%%%%%%%%%%%
\section{Introduction}
Biosensors play an important role in metrological studies of biological and chemical elements in clinical health, agricultural, and environmental applications \cite{MolinaFernndez2019, Mehrotra2016, Bhalla2016, Haleem2021, Alemdar2024}. Detection of ultra-low concentration solutes is a challenging problem with the additional requirements of low-cost, high sensitivity, high-speed testing, portable sensor device, and label-free detections on the sensing scheme \cite{Brandenburg1994, Liu1992, Ymeti2006}. Clinical applications demand the detection of low concentrations (c) of analytes (biomolecules, chemicals, particles, etc.) in the range of $\mu$g/ml, corresponding to refractive index ($\text{n}_{\text{sol}}$) changes in the order of $10^{-7} \,\text{RIU}\, (\delta \text{n}_{\text{sol}}/ \delta \text{c} \sim 10^{-1} \,\text{RIU}/(\text{g/ml}))$, where RIU is refractive index unit \cite{MolinaFernndez2019, Estevez2011, Tan2015}.  Optical biosensors provide robust ultra-high sensitivity technology in the measurement of solute concentrations in $\mu$g/ml order \cite{Estevez2011, Leuermann2019}. Optical or photonic devices that manipulate the light waves, such as interferometers, achieve phase shifts, and intensity variations are the fundamental building blocks in many applications like sensors, modulators, and optical switches \cite{Butt2023, Sinatkas2021, Chen2023}. Photonic biosensors can integrate all biosensing components onto a CMOS-compatible chip, enabling multiplexing with compact electronic readouts and offering the advantage of a system with smaller noise levels \cite{Altug2022}.
\\

\noindent
Various types of sensors are capable of sensing tiny concentrations of solute or analyte. Surface plasmon resonance (SPR) and microcavity sensors depend on resonance effects, which are sensitive to environmental fluctuations and necessitate precise resonance measurements \cite{Miyazaki2017, Taniguchi2016}. Photonic crystal sensors consist of structures spatially arranged periodic dielectric materials that uniquely interact with light and are sensitive to the changes in environment refractive index\cite{Inan2017, Gowdhami2022}. Inverse design methods use computational algorithms to optimize photonic structures for specific functionalities, such as refractive index sensing, often resulting in unconventional geometries tailored for maximum performance \cite{Singh2021, Chung2021, DidariBader2024, Chung2022}. However, photonic waveguides, while perhaps less optimized than inverse-designed structures, offer several practical advantages. Waveguides are based on well-established fabrication techniques, making them more straightforward to manufacture with high yield and reproducibility. They also tend to be more robust to fabrication imperfections, whereas inverse-designed structures can be susceptible to small deviations from the ideal design, potentially leading to performance degradation. Inverse-design-based sensors necessitate heavy computational optimizations and high-resolution electron-beam lithography (EBL) methods in fabrication. In contrast, photonic waveguides can be easily simulated, optimized, and fabricated using the photolithography method, and the measurements are simpler to execute, as reported here.
\\

\noindent
Photonic biosensors, employing configurations such as Young's interferometer (YI) in photonic waveguides do not require the EBL method in fabrication and enable accurate measurements of solute concentrations. However, external temperature and pressure, laser wavelength, linewidth and stability, low precision or noisy detection schemes, and fabrication errors can increase total noise in the sensor signal, therefore reduce the sensor's sensitivity and limit detection capabilities. The lowest concentrations that can be reliably detected in the employed sensing scheme are commonly referred to as the limit of detection (LoD) and depend both on the sensor's sensitivity and the read-out system's noise floor. Common ways to increase the sensitivity of sensors are by increasing the interaction length, optimizing the sensing arm waveguide, reducing the instrument noise floor in the system, and making high precision measurements of the phase shift \cite{Ymeti2006, Martens2019, MolinaFernndez2019}. New techniques to counter such effects have also been demonstrated, such as temperature-independent MZI-based biosensors and coherent interferometric sensors \cite{MolinaFernndez2019, Luo2022, GonzlezGuerrero2016}. 
\\

\noindent
Thin film silicon nitride ($\text{Si}_{3}\text{N}_{4}$) offers a prominent platform for integrated photonic circuits due to their low propagation loss in visible-near infrared wavelengths, high refractive index, low-cost and compatibility with CMOS fabrication processes \cite{Alajlan2020, Xiang2022}. MZI-based sensors require its operation near the quadrature condition for high sensitivity measurement of phase differences, but Young's interferometer is applicable for arbitrary phase shifts. In addition, the maturity of the CMOS industry enables low-cost instrumentation of image-based sensing, whereas MZI-based sensors still require an expensive photodetection scheme. Here, we implement Young's interferometer in  $\text{Si}_{3}\text{N}_{4}$ photonic waveguide platform, where $\text{Si}_{3}\text{N}_{4}$ is a core waveguide material and silicon dioxide ($\text{SiO}_{2}$) serves as cladding on silicon (Si) substrate. We confine the light in the waveguide's fundamental transverse electric (TE) mode and build a photonic sensor. We demonstrate its capability of sensing refractive index change through fringe shifts with different concentrations of glucose introduced to the sensing arm of the interferometer. Our sensor's sensitivity performs better than that of the works reported by Wang et al. \cite{Wang2012}, Zhou et al. \cite{Zhou2018}, and Wong et al. \cite{Wong2019}. 

\begin{figure}[ht!]
\centering\includegraphics[width=12cm]{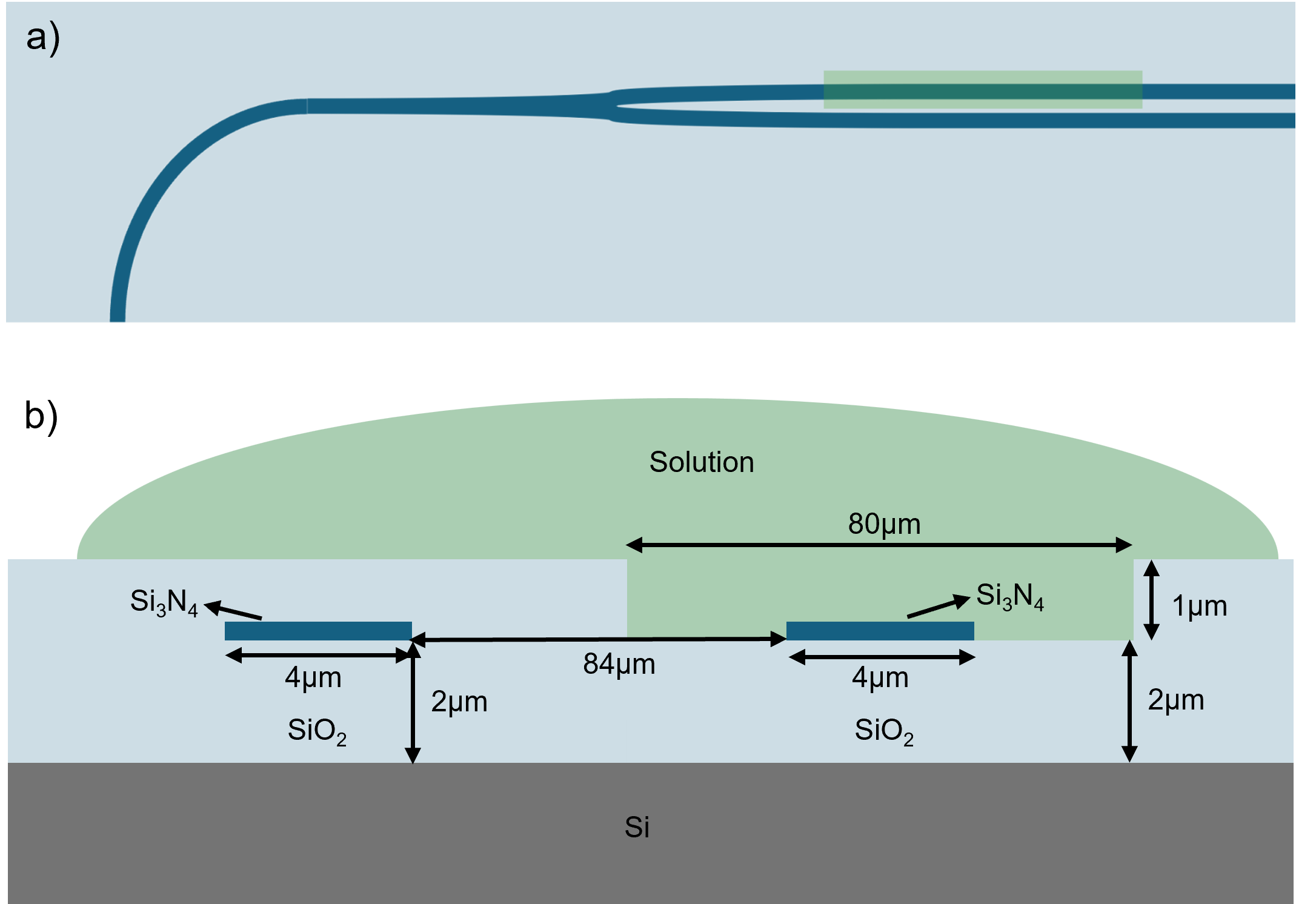}
\caption{Schematic diagrams: (a) Top view of Young's interferometer implemented in silicon nitride photonic waveguide platform, a waveguide splits into two waveguides using a Y-splitter, one waveguide is the sensing arm (sensing window is shaded as green) and other acts as the reference arm. (b) Cross section of the chip in the sensing region, marked dimensions are used in chip fabrications in this work. Note: The schematics are not to scale.}
\label{yi_schematic_diagram}
\end{figure}

\section{Design and Simulations}

The operating principle of waveguide sensors is based on the evanescent electromagnetic (EM) wave of confined mode in the waveguide interacting with the environment in the vicinity of the waveguide surface. The propagation of EM wave is susceptible to the environment refractive index due to the mode fields' presence outside the waveguide, which causes changes in the effective index of mode due to small perturbations in the environment. This interaction allows us to develop a technology to measure tiny changes in analyte concentration in the aqueous solution. The change in phase of an EM wave is detected by interfering it with another non-interacting beam in the interferometer \cite{Watchi2018}. 
\\

\noindent
Waveguide-based YI splits a waveguide into two waveguides via a 50:50 Y-splitter; one waveguide is the sensing arm, and the other acts as the reference arm. The 50:50 Y-splitter divides the incoming guided light intensity equally  into each waveguide, as well as prevents optical losses and conversion to other modes. The chip configuration of waveguide-based YI is shown in Fig. \ref{yi_schematic_diagram}. The two confined modes in the sensing and reference waveguides come out of the chip end, form diverged beams, and interfere in the free space. The interference pattern is captured by a CMOS camera. The reference waveguide is buried in $\text{SiO}_{2}$, and the sensing arm is open to the aqueous medium as shown in Fig. \ref{yi_schematic_diagram}(b). The propagating mode in the sensing arm interacts with the solution, causing an additional phase shift with respect to the mode propagating in the reference arm. When the solution's refractive index changes, the phase of the propagating mode in the reference arm remains constant, while the phase of the propagating mode in the sensing arm changes. This phase shift of sensing arm mode is given by, $\delta \phi = 2\pi L \Delta n_{\text{eff}}/\lambda$, where $L$ is the length of sensing window, $\Delta n_{\text{eff}}$ is the change in the effective index of sensing arm waveguide mode, $\lambda$ is the wavelength of the light. The ratio of the variation in the effective index of the sensing arm mode to the variation in the solution's refractive index is defined as the bulk sensitivity of the sensing arm waveguide or the sensor, $\text{S}_{\text{wg}}$,

\begin{equation} \label{eq:1}
\text{S}_{\text{wg}} = \frac{\delta \text{n}_{\text{eff}}}{\delta \text{n}_{\text{sol}} }
\end{equation}

\noindent
where $\delta \text{n}_{\text{eff}}$ is change in effective index of mode and $\delta \text{n}_{\text{sol}}$ is change in solution's refractive index.
\\ 

\begin{figure}[ht!]
\centering\includegraphics[width=12cm]{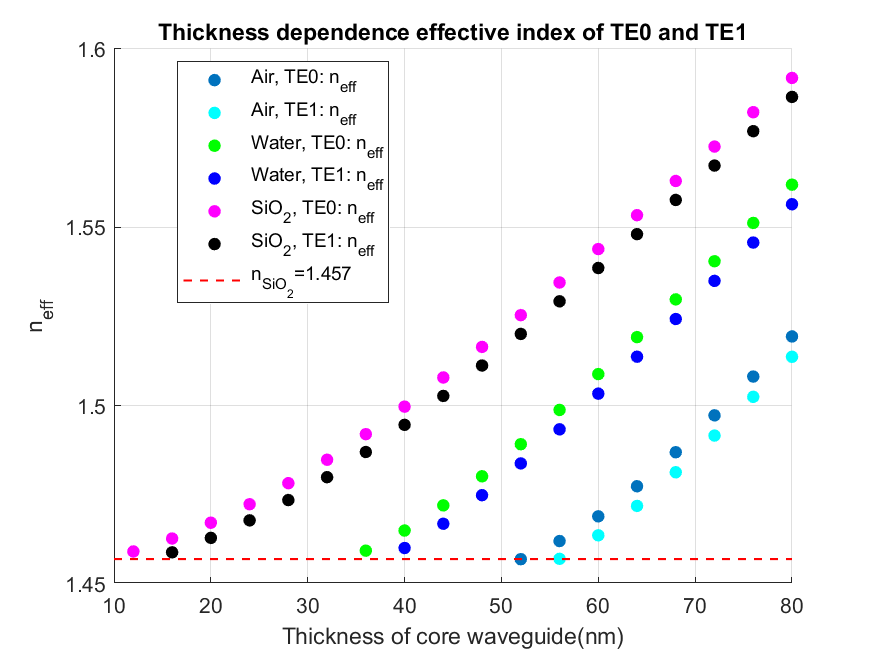}
\caption{The effective indices of the TE0 and TE1 modes at 633 nm for different thicknesses and different surrounding materials (air, water, and $\text{SiO}_{2}$) on top of a $4\, \mu$m wide $\text{Si}_{3}\text{N}_{4}$ core waveguide. For all cases, the $\text{Si}_{3}\text{N}_{4}$ core waveguide sits on top of  $2 \, \mu$m thick $\text{SiO}_{2}$ cladding on Si substrate as shown in Fig. \ref{yi_schematic_diagram}. The black and purple dotted curve corresponds to the $\text{Si}_{3}\text{N}_{4}$ core waveguide covered by $1\, \mu$m in $\text{SiO}_{2}$; green and royal blue dotted curve corresponds to water on top of the $\text{Si}_{3}\text{N}_{4}$ core waveguide; and light blue and cyan dotted curve corresponds to air on top of $\text{Si}_{3}\text{N}_{4}$ core waveguide. The red dotted line is the refractive index of $\text{SiO}_{2}$ cladding.}
\label{eff_modes_plot}
\end{figure}

\begin{figure}[ht!]
\centering\includegraphics[width=12cm]{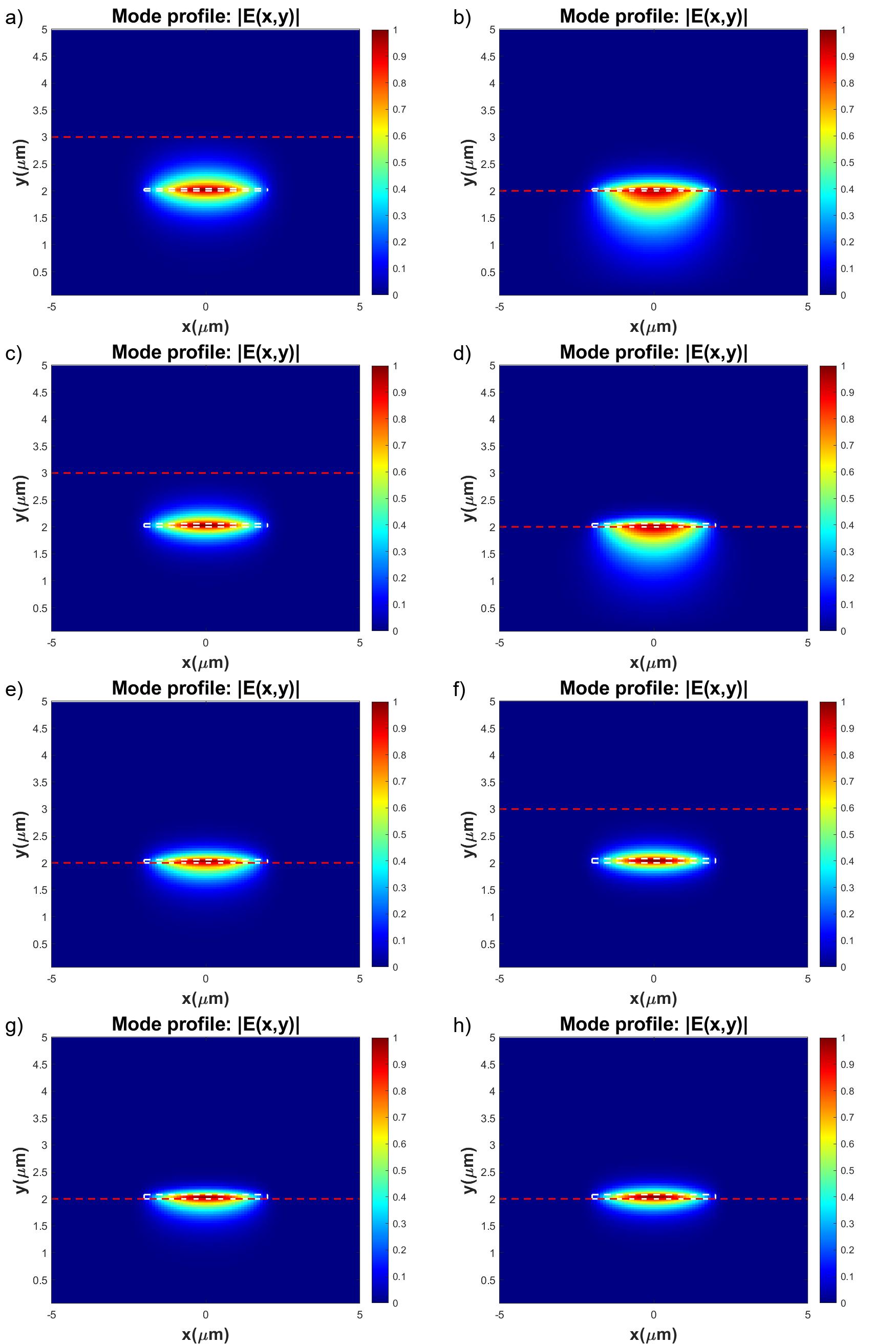}
\caption{Mode profiles of TE0 modes confined in different thicknesses and surrounding media on top of the $4\, \mu$m wide $\text{Si}_{3}\text{N}_{4}$ core waveguide: (a) 40 nm $\text{Si}_{3}\text{N}_{4}$ with $\text{1} \, \mu$m $\text{SiO}_{2}$ on top, (b) 40 nm $\text{Si}_{3}\text{N}_{4}$ with water  on top, (c) 60 nm $\text{Si}_{3}\text{N}_{4}$ with $\text{1} \, \mu$m $\text{SiO}_{2}$ on top, (d) 60 nm $\text{Si}_{3}\text{N}_{4}$ with air on top, (e) 60 nm $\text{Si}_{3}\text{N}_{4}$ with water  on top, (f) 80 nm $\text{Si}_{3}\text{N}_{4}$ with $\text{1} \, \mu$m $\text{SiO}_{2}$ on top, (g) 80 nm $\text{Si}_{3}\text{N}_{4}$ with air on top and (h) 80 nm $\text{Si}_{3}\text{N}_{4}$ with water on top of core waveguide. The white and red dotted lines show the core waveguide and $\text{SiO}_{2}$ interface with air or water, respectively.}
\label{mode_prolifes_plot}
\end{figure}

\noindent
The effective indices of modes confined in the waveguides are obtained using numerical solutions of Maxwell's equations. We use finite-difference time-domain (FDTD) analysis to estimate the effective indices of modes and mode profiles. At the operating wavelength of our laser, 633 nm, the refractive index of $\text{SiO}_{2}$, $\text{Si}_{3}\text{N}_{4}$ and pure water are taken as 1.457, 2.01 and 1.33, respectively. Fig. \ref{eff_modes_plot} shows the effective index of TE modes dependence on  $\text{Si}_{3}\text{N}_{4}$ core waveguide thickness for buried waveguide in $\text{SiO}_{2}$ and sensing arm waveguide in air and water. Since the cladding $\text{SiO}_{2}$ index is 1.457, any mode with an effective index lower than the cladding index would not be guided in the core waveguide, and as the core waveguide's thickness increases, the modes' effective indices increase. The plot also shows that as the index of surrounding material on top of the core waveguide decreases, a larger thickness of the core waveguide is required for the existence of the guided modes. For instance, 40 nm thick 4 $\mu$m wide $\text{Si}_{3}\text{N}_{4}$ core waveguide doesn't support a guided mode with surrounding material air on top, thus this configuration is not plotted in Fig. \ref{mode_prolifes_plot}. To confine the mode in sensing arm waveguide covered with water, a thicker core waveguide is needed compared to the thickness required for single-mode (SM) operation in the buried waveguide. This is because the effective indices of the confined modes in the buried waveguide are higher than those in the sensing arm waveguide covered with water.

\begin{figure}[ht!]
\centering\includegraphics[width=12cm]{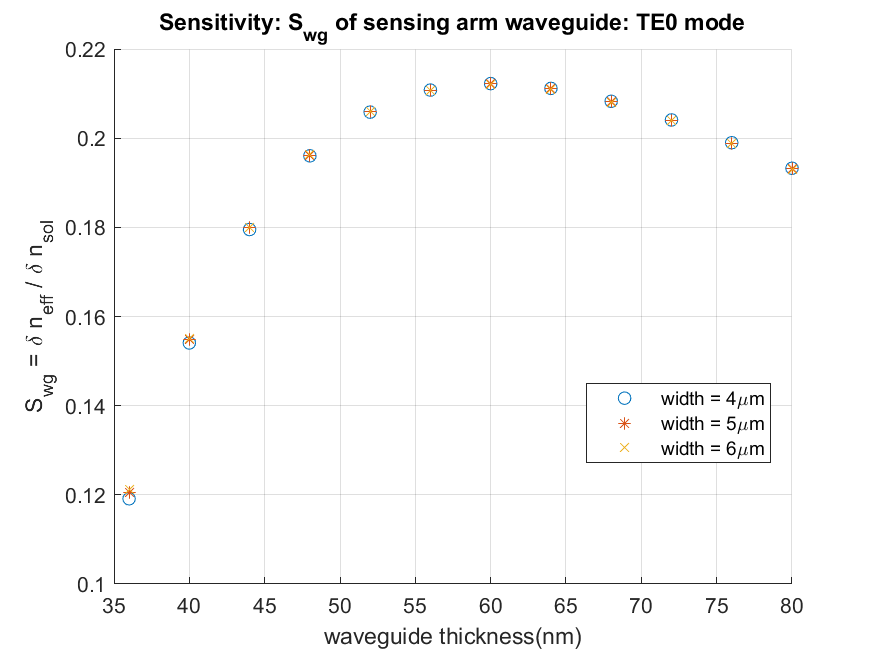}
\caption{Sensitivity of fundamental TE (TE0) mode confined in sensing arm waveguide for different thicknesses and width of $\text{Si}_{3}\text{N}_{4}$ core waveguide.}
\label{sensitivity_plot}
\end{figure}

\begin{figure}[ht!]
\centering\includegraphics[width=12cm]{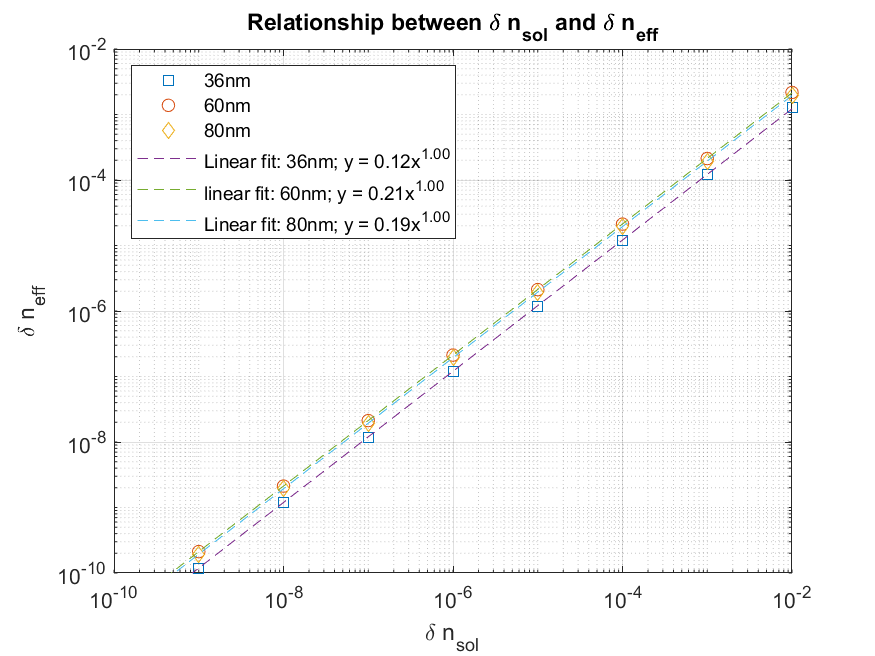}
\caption{The relationship between change in the effective index of TE0 mode of sensing arm waveguide and solution bulk index change for 4$\, \mu$m wide and 36 nm, 60 nm, and 80 nm thick core waveguide as the configuration used in Fig. \ref{sensitivity_plot}. The curves are fitted with a linear function (y = mx), where the slope m is $\text{S}_{\text{wg}}$ plotted in Fig. \ref{sensitivity_plot}.}
\label{sensitivity_1}
\end{figure}

\clearpage
\noindent
Fig. \ref{mode_prolifes_plot} shows the fundamental mode profiles confined in three different thicknesses: 40 nm, 60 nm, and 80 nm, with 4 $\mu$m wide core waveguide covered by 1 $\mu$m $\text{SiO}_{2}$, bulk water or air, respectively. First, the modes are well confined within the 1 $\mu$m $\text{SiO}_{2}$ at all core thicknesses, and they are quite symmetric with respect to the horizontal symmetry axis of the core. The mode profile is asymmetric for sensing arm waveguide due to asymmetry of refractive indices of water and $\text{SiO}_{2}$ as shown in Fig. \ref{mode_prolifes_plot}. The fraction of mode fields in solution, i.e., the portion above the red dotted line indicating the interface, decreases for thinner waveguides, thus in order to increase the sensitivity, the core waveguide's thickness must be kept larger than SM criteria in general \cite{Milvich2018}. The sensitivity dependence on the core waveguide thickness is shown in Fig. \ref{sensitivity_plot}. The relationship between $\delta \text{n}_{\text{sol}}$ and $\delta \text{n}_{\text{eff}}$ is shown in Fig. \ref{sensitivity_1} for three out of 36 data points in Fig. \ref{sensitivity_plot}, corresponding to 36 nm, 60 nm and 80 nm thick and 4 $\mu$m wide core waveguide. Fig. \ref{sensitivity_plot} shows that the thickness of the core waveguide for maximum sensitivity is larger than the thickness satisfying the SM condition; this can be seen from Fig. \ref{eff_modes_plot} where TE1 mode starts to exist. Furthermore, sensitivity decreases for the larger thickness of the core waveguide than the thickness for the maximum sensitivity due to the increase of mode confinement in the core waveguide and the decrease in mode field overlap with water or solution.
\\

\begin{figure}[ht!]
\centering\includegraphics[width=12cm]{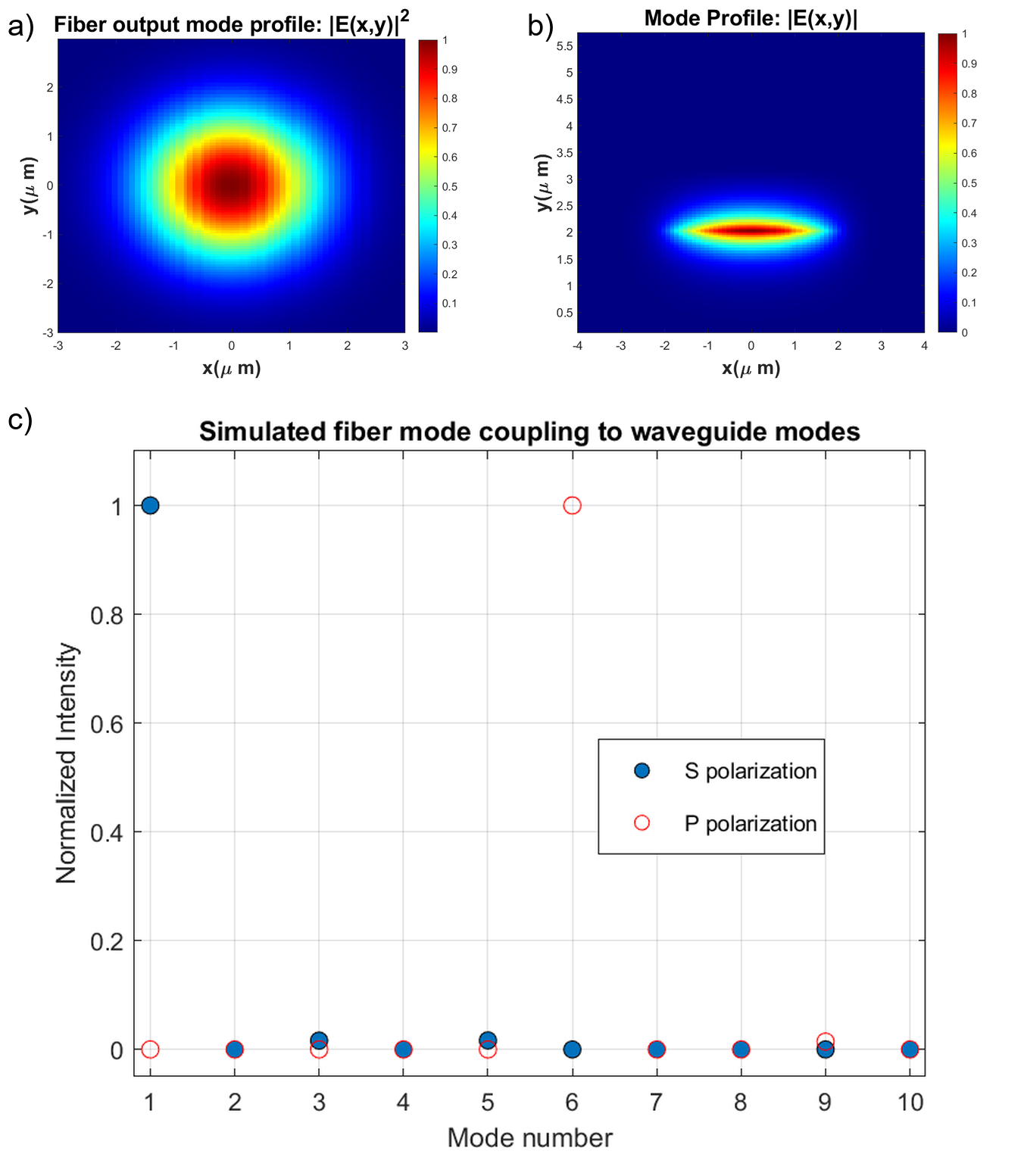}
\caption{Simulated coupling of TEM00 mode of SM fiber to modes confined in 54 nm thick, $4 \,\mu$m wide $\text{Si}_{3}\text{N}_{4}$ core waveguide buried in $\text{SiO}_{2}$, the configuration is discussed in Fig. \ref{yi_schematic_diagram}(b). (a) The intensity profile of TEM00 mode of SM fiber. (b) Mode profile of TE0 mode confined in buried $\text{Si}_{3}\text{N}_{4}$ core waveguide. (c) Normalized intensity of mode overlap of TEM00 mode of SM fiber to different modes (mode number on the x-axis) confined in the core waveguide at the fiber-chip coupling interface. Note: images of mode profiles of higher modes confined in the buried waveguide are not included, but coupling from SM fiber mode to all the confined modes in the buried waveguide is simulated and plotted in (c).}
\label{fiber_mode_coupling_Sin_wg_SiO2}
\end{figure}

\begin{figure}[ht!]
\centering\includegraphics[width=12cm]{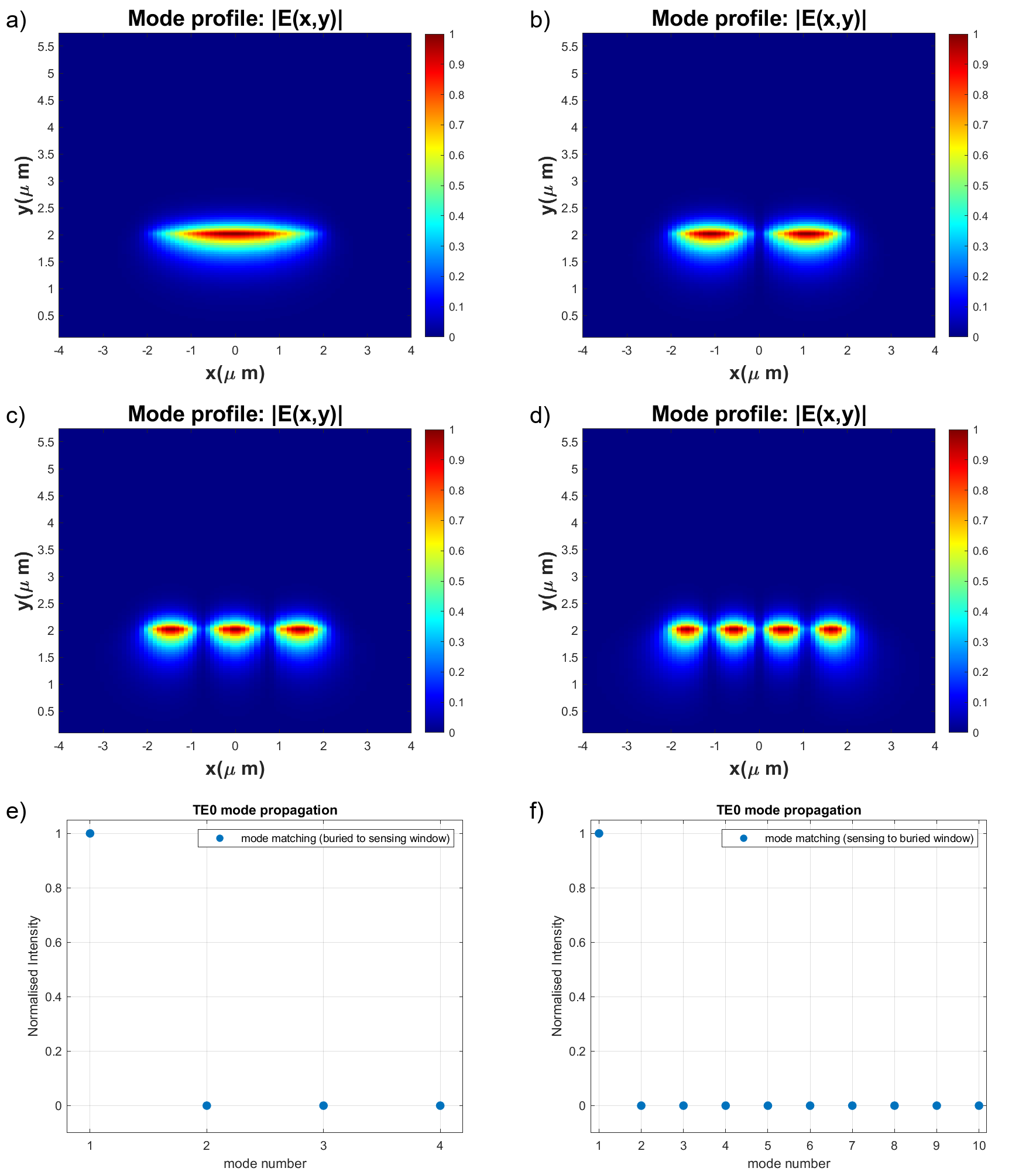}
\caption{Simulated coupling of TE0 modes at the interface of sensing window as configuration shown in Fig. \ref{yi_schematic_diagram} with 54 nm thick, $4\, \mu$m wide $\text{Si}_{3}\text{N}_{4}$ core waveguide. (a), (b), (c), and (d) are TE0, TE1, TE2, and TE3 confined modes in the sensing arm waveguide. Normalized intensity of mode overlap of TE0 mode to different modes (mode-number on the x-axis): (e) transition from buried waveguide to sensing arm waveguide and (f) transition from sensing arm waveguide to buried waveguide (the same 10 modes on the x-axis of Fig. \ref{fiber_mode_coupling_Sin_wg_SiO2}(c)). }
\label{mode_matching_sensing}
\end{figure}

\noindent 
The design of the curved waveguide and interferometer implemented on the chip is adjusted such that the output from the chip is free of scattered light and unguided cladding modes that may occur at the input fiber-chip interface and sensing window area as shown in Fig. \ref{yi_schematic_diagram}. For a demonstration of the sensor's sensing application, we utilize the following dimensions for our waveguide-based interferometer: the width of $\text{Si}_{3}\text{N}_{4}$ core waveguide is $4 \, \mu$m. The gap between two waveguides of the interferometer is $84 \, \mu$m (edge to edge) as shown in Fig. \ref{yi_schematic_diagram}(b). The sensing window measures 12 mm long, $80 \, \mu$m wide, centered on the sensing arm waveguide and $1 \, \mu$m deep. We use 54 nm thick, 4 $\mu$m wide $\text{Si}_{3}\text{N}_{4}$ core waveguide in the demonstration experiment as discussed in the next sections. Since the coupling of light from SM fiber to waveguide (experimental setup is described in section 4) is mostly into fundamental TE mode of the core waveguide as shown in Fig. \ref{fiber_mode_coupling_Sin_wg_SiO2} and only TE0 mode propagates as depicted in Fig. \ref{mode_matching_sensing}, the higher modes are not excited in the sensor and do not play any role in our sensor measurements. The simulated $\text{n}_{\text{eff}}$ of TM0 mode for 54 nm thick, 4 $\mu$m wide $\text{Si}_{3}\text{N}_{4}$ sensing arm waveguide is 1.4556 which is below $\text{SiO}_{\text{2}}$ cladding index 1.457 at 633 nm, therefore TM0 mode is not confined in sensing arm waveguide and do not play a role in our experiment. 
\\ 

\noindent
In Fig. \ref{yi_schematic_diagram}(a), light propagates from the buried waveguide into the sensing window and couples back into the buried waveguide. Fig. \ref{mode_matching_sensing} shows a total of four guided modes in the sensing window for 54 nm thick, 4 $\mu$m wide $\text{Si}_{3}\text{N}_{4}$ core waveguide, including a fundamental and three higher TE modes. The propagating TE0 mode from the buried waveguide does not couple to any of those three higher modes in the sensing window. Similarly, the propagating TE0 mode in the sensing window does not couple to any higher mode confined in the buried waveguide. In general, it is advised to use the adiabatic waveguide design at the transition regions of the sensing window (buried waveguide to sensing arm waveguide and sensing arm waveguide to buried waveguide) to improve the mode conversion from buried waveguide to sensing arm waveguide and vice versa. Typically, a triangular shape with dimensions estimated by suitable simulations can be implemented at both ends of the sensing window to reduce light scattering and coupling to any mode other than the fundamental TE mode. However, this adiabatic design was not used in the demonstrated chip presented here.

\clearpage

\section{Fabrication of Photonic Sensor Chip}
\begin{figure}[ht!]
\centering\includegraphics[width=12cm]{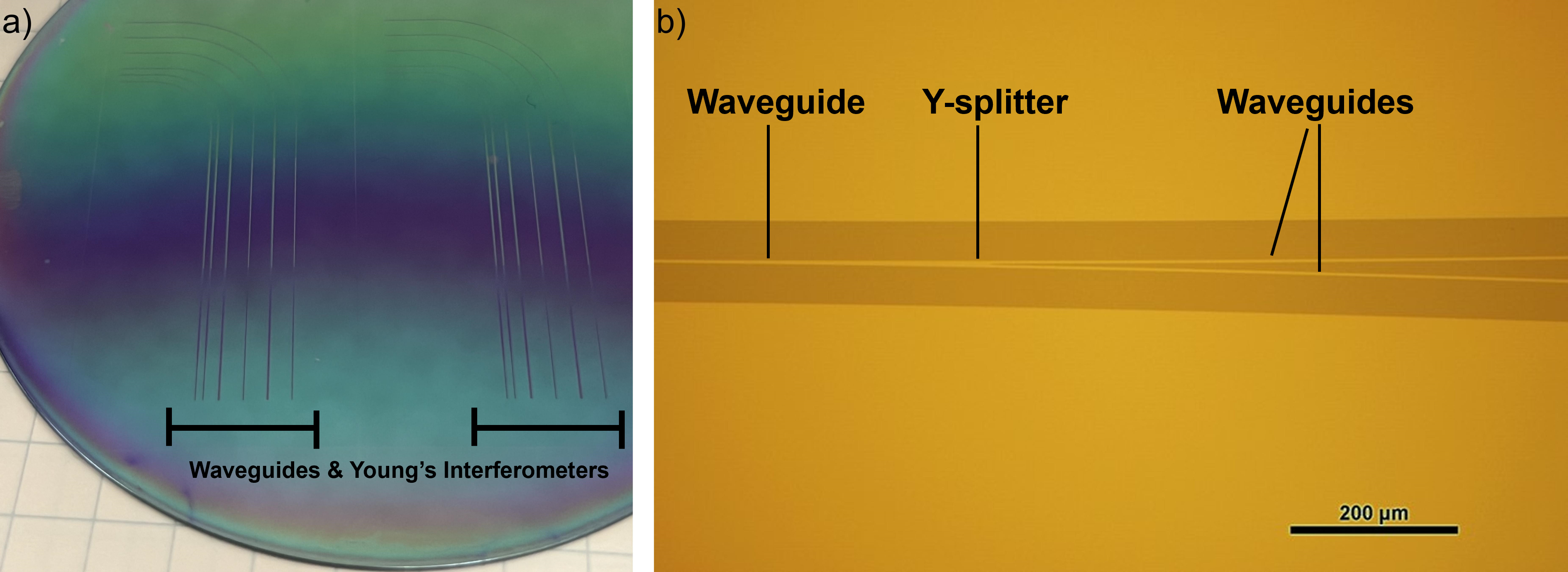}
\caption{(a) Fabricated 3-inch photonic chip with sets of waveguides and YI sensors. (b) Optical microscope image of Y-splitter. See Fig. \ref{yi_schematic_diagram} (a) for the schematic diagram.}
\label{fabricated_chip}
\end{figure}

\begin{figure}[ht!]
\centering\includegraphics[width=12cm]{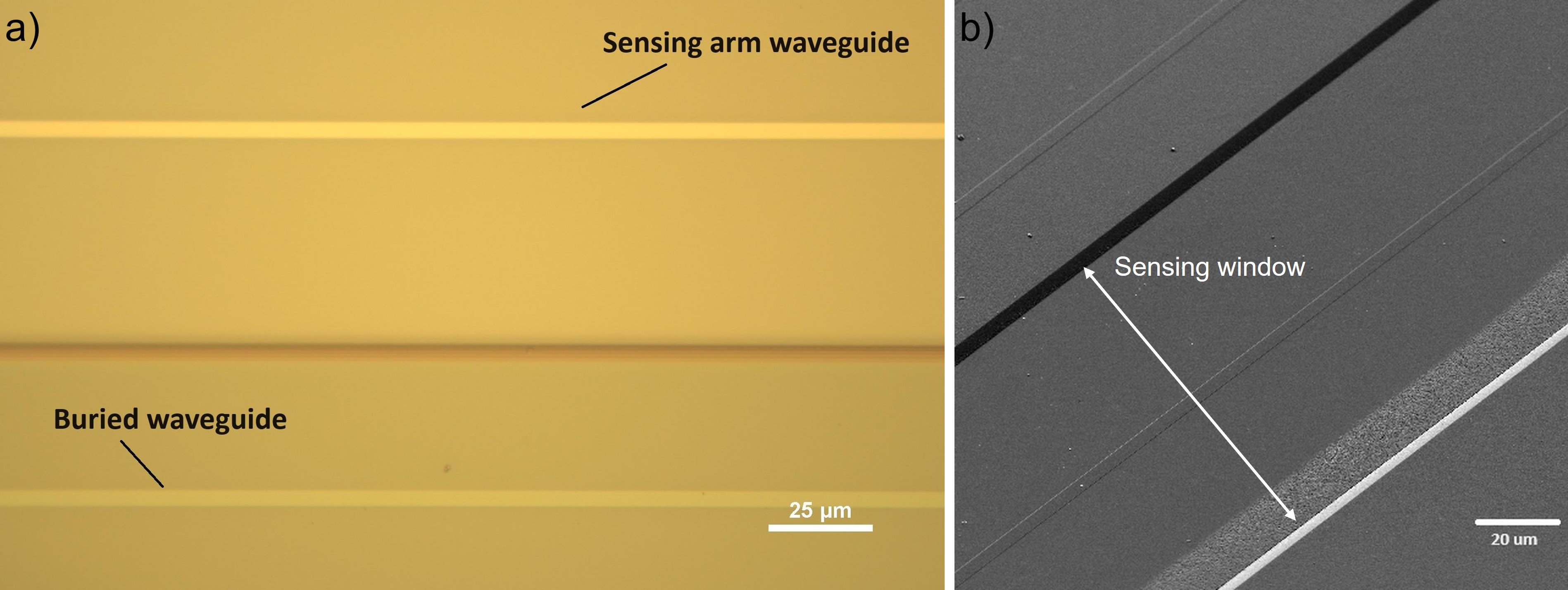}
\caption{(a) Optical microscope image of chip showing $\text{Si}_{3}\text{N}_{4}$ sensing arm waveguide and buried waveguide. (b) Electron microscope image of sensing window region with sample tilted at 50 degrees.}
\label{sensing_region_images}
\end{figure}

The fabrication of a photonic sensor chip begins with a 3-inch silicon (Si) wafer. A layer of silicon dioxide ($\text{SiO}_{2}$) with a thickness of $2\, \mu$m is thermally grown on the wafer's surface to provide a base isolation layer. Next, a thin layer of $\text{Si}_{3}\text{N}_{4}$ approximately 73 nm is deposited using the Low-Pressure Chemical Vapor Deposition (LPCVD) technique. This $\text{Si}_{3}\text{N}_{4}$ layer forms the core of the waveguide. The deposited $\text{Si}_{3}\text{N}_{4}$ layer is then etched down to the desired waveguide thickness using a plasma etching process, i.e., 54 nm thick $\text{Si}_{3}\text{N}_{4}$ in the demonstration experiment. The wafer is baked on a hot plate for 10 minutes at 160 $^{\circ}$C to dehumidify the surface of the wafer. A positive photoresist, S1818, is then spin-coated onto the wafer at 3500 revolutions per minute (rpm) for 40 seconds, resulting in a resist layer of about $2\, \mu$m thick. This resist layer acts as a mask during the following patterning process. The resist-coated wafer undergoes a soft bake on a hot plate at 115 $^{\circ}$C for 2 min to solidify the resist and improve its patterning properties. After cooling down to room temperature, the wafer is then transferred to the photo-lithography system (Heidelberg MLA150 Maskless) to pattern the interferometer and waveguide designs on photoresist with ultraviolet (UV) light exposure.  The exposure dose is set to $160\, \text{mJ}/\text{cm}^{2}$ to achieve optimal resist patterning. The exposed resist is then developed in a developer solution, MF-319, for 90 seconds. This development process removes the unexposed regions of the resist, leaving behind the designed waveguide pattern. The patterned resist layer acts as a mask for the subsequent reactive-ion etching (RIE) process, which precisely etches away uncovered $\text{Si}_{3}\text{N}_{4}$ layer. Finally, the remaining resist mask is removed using acetone, completing the fabrication of the core waveguide structure. After that, $1\, \mu$m thick $\text{SiO}_{2}$ is deposited on the fabricated chip using the PECVD technique. To create a sensing window on the sensing arm waveguide, positive photoresist S1818 is spin-coated onto the wafer at 3500 rpm for 40 seconds to form a resist layer around $2\, \mu$m, then soft baking at 115 $^{\circ}$C for 2 min is performed. The wafer is then transferred to EVG 610 Double-sided Mask Aligner photo-lithography system to align to markers fabricated waveguides during the previous lithography step and pattern the sensing windows on photoresist with UV exposure dosage $160\,\text{mJ}/\text{cm}^{2}$. The pattern is developed in the MF-319 developer solution for about 90 seconds. After development, a wet etching method, Buffered oxide etch (BOE) chemical is used for 4 min to etch $\text{SiO}_{2}$ from patterned sensing areas down to $\text{Si}_{3}\text{N}_{4}$ core waveguide. Following the etching, the resist mask is lifted off with acetone, leaving behind the completed photonic waveguide with a defined sensing region ready for further characterization and integration as shown in Fig. \ref{fabricated_chip} and \ref{sensing_region_images}. The chip is then cleaved with a diamond cutter and gentle pressure by a clip to propagate the cleave cut.

\section{Experimental Setup}
\begin{figure}[ht!]
\centering\includegraphics[width=12cm]{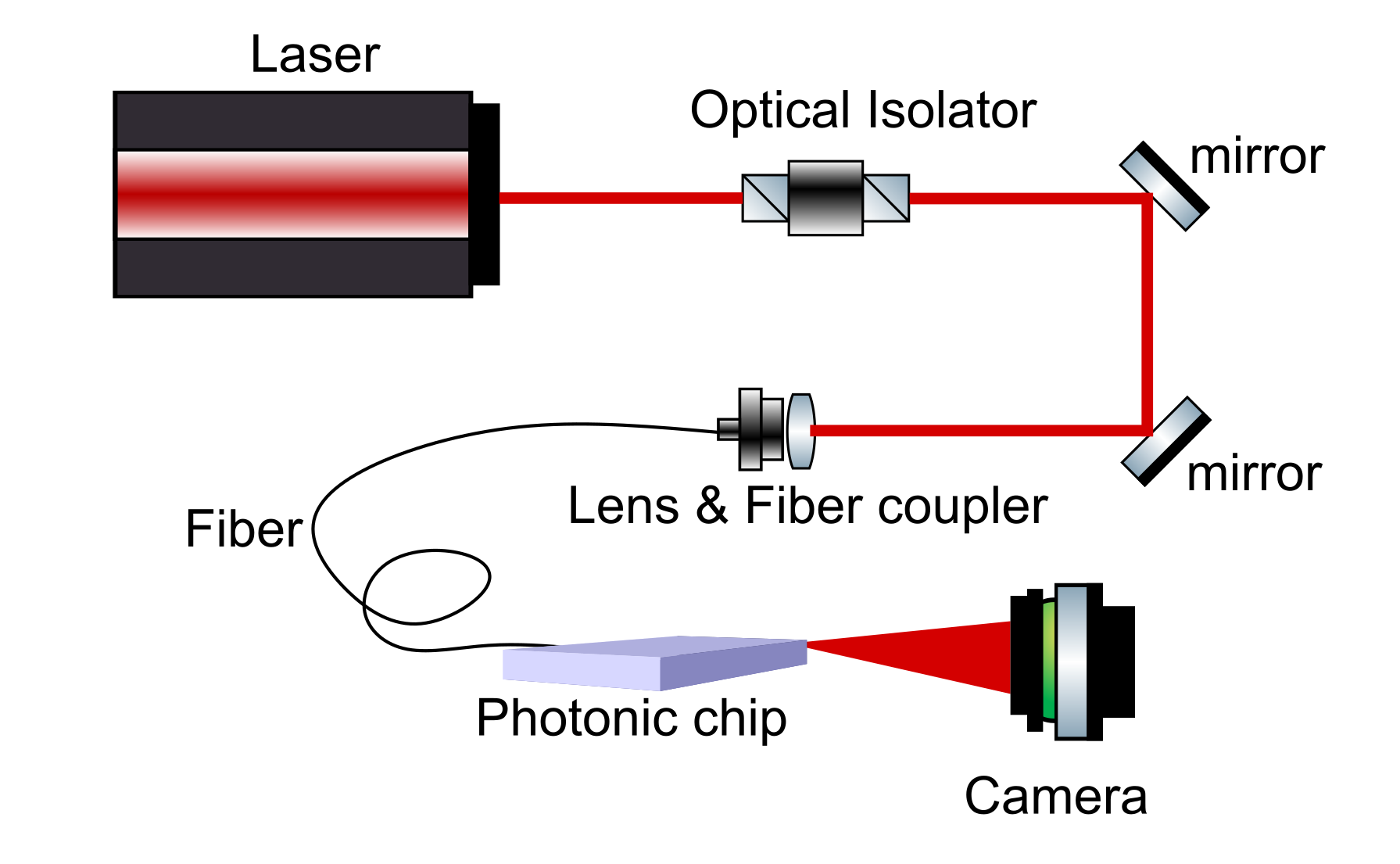}
\caption{Schematic diagram of the optical setup to test photonic sensor chips.}
\label{schematic_diagram}
\end{figure}

The schematic diagram of the photonic chip characterization experiment is shown in Fig. \ref{schematic_diagram}. A continuous-wave (CW) laser (Thorlabs HRS015B)  operating at a wavelength of 633 nm is launched to an optical isolator (Newport ISO-04-650-MP) to minimize any reflection back to the laser and is then coupled into a SM fiber (Thorlabs P1-630Y-FC-2) with a core diameter of approximately 4 $\mu$m.  One end of the SM fiber is cleaved at a 0-degree angle to have a perfectly Gaussian fundamental Transverse EM (TEM) mode at the output. A photonic sensor chip is placed under an optical microscope equipped with $5\times$ and $20\times$ objective lenses. The cleaved fiber end is then launched from a control stage towards the photonic chip and butt-coupled to the $\text{Si}_{3}\text{N}_{4}$ core waveguide to achieve the coupling of light. The output from the chip is captured by a CMOS camera and monitored on a computer with a Labview application, allowing us to visualize the interference pattern and apply further data analysis. 

\section{Results and Discussions}
\begin{figure}[ht!]
\centering\includegraphics[width=12cm]{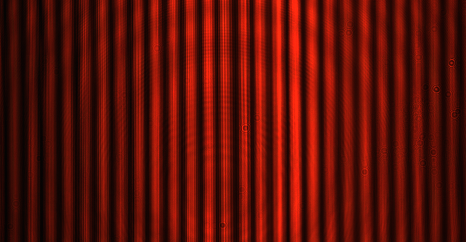}
\caption{Interference pattern captured in a CMOS camera from photonic waveguide-based YI sensor chip output. The sensing window region is filled with glucose solution.}
\label{yi_output_water}
\end{figure}

\begin{figure}[ht!]
\centering\includegraphics[width=12cm]{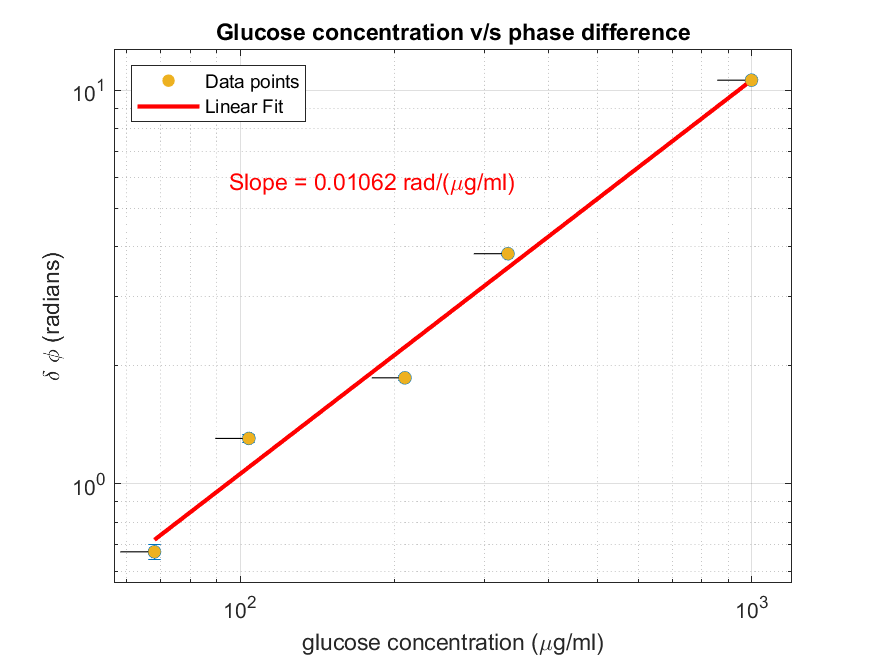}
\caption{Plot of measured phase differences with varied glucose concentration solution with pure water as the reference.}
\label{result_plot}
\end{figure}

For the glucose solution concentration measurement, we fabricate a photonic chip with 54 nm thick $\text{Si}_{3}\text{N}_{4}$ core waveguide, which is close to the optimum thickness for sensitivity as discussed in Fig. \ref{sensitivity_plot}. The simulated coupling of TE0 waveguide mode to higher modes at sensing window interfaces is shown in Fig. \ref{mode_matching_sensing}, illustrating our sensor's operation in fundamental TE mode. Drops of solution are placed on the sensing window using a syringe from prepared glucose (D-glucose, Sigma-Aldrich) solutions to fill the sensing area of the sensor device. We wait for 5 min after putting solution for each measurement to guarantee the equilibrium state of the interaction of solution with sensing arm waveguide and monitor fringe shifts with live setup of camera output on Labview application with only less than 100 ms integration time for each output image captured. Interfered beam output from the chip end is a diverging beam, and the camera collects a small region of light output for about 80 ms integration time, as shown in Fig. \ref{yi_output_water}. The camera is placed away from the chip to capture only 4-5 fringes on the camera to get enough pixels of data points per fringe along the fringe shift axis (x-axis). The fringes produced by our photonic chip have visibility ($(\text{I}_{\text{max}}- \text{I}_{\text{min}})/(\text{I}_{\text{max}}+ \text{I}_{\text{min}})$) more than 0.75 in all measurements. We use the $A\sin(\omega x +\phi) + b$ function to fit the fringe data along the fringe shift axis (x-axis) to estimate the output phase. We fit the data with a fitting function using the Python scipy module and extract the fit parameters. Based on the data fittings of multiple fringes, the estimated phases' standard deviation is $\sigma_\phi$  $\sim 0.03$ radians. Phase differences, $\delta \phi$, are calculated for different glucose conc. solutions using the pure water solution as a reference, i.e., $\delta \phi = \phi_{conc.} - \phi_{0}$, where $\phi_{0}$  and $\phi_{conc.}$ are estimated phases in radians for pure water and glucose conc. solutions respectively. The measured values $\delta \phi$ vs conc. of glucose are plotted in Fig. \ref{result_plot}, and the data fits a linear function relation between conc. and phase differences.  Based on the work of Tan et al. \cite{Tan2015}, at room temperature $\sim \text{300}$ K, $\delta \text{n}_{\text{sol}}/ \delta \text{c}$ for glucose conc. solution is $\sim 1.56 \times 10^{-1} \,\text{RIU}/(\text{g/ml})$. The smallest glucose conc. used in this experiment is 68 $\mu g/ml$, which corresponds to bulk refractive index change of $\sim 1.06 \times 10^{-5} \,\text{RIU}$ from pure water.

\noindent
To determine the number of molecules around the waveguide contributing to phase shift in interference, first, we estimate the effective mode area \cite{Agrawal2019} using the following equation:

\begin{equation} \label{eq:2}
\text{A}_\text{eff} = \frac{\iint \text{n}^{2}\text{(x,y)}|\text{E(x,y)}|^{2} \,\text{dx}\,\text{dy}}{\max[\text{n}^{2}\text{(x,y)}|\text{E(x,y)}|^{2}]} ,
\end{equation}

\noindent where, $\text{A}_\text{eff}$ is the effective mode area \cite{Agrawal2019}, $|\text{E(x,y)}|$ is the magnitude of electric field and $\text{n(x,y)}$ is the refractive index. For a mode spans waveguide and solution, the effective mode area $\text{A}_\text{eff, mode}$ is 
\begin{equation} \label{eq:3}
\text{A}_\text{eff, mode} = \frac{\iint_\text{mode} \text{n}^{2}\text{(x,y)}|\text{E(x,y)}|^{2} \,\text{dx}\,\text{dy}}{\max[\text{n}^{2}\text{(x,y)}|\text{E(x,y)}|^{2}]_\text{mode}} ,
\end{equation}
while the effective mode area in the solution is
\begin{equation} \label{eq:4}
\text{A}_\text{eff, solution} = \frac{\iint_\text{sol} \text{n}^{2}\text{(x,y)}|\text{E(x,y)}|^{2} \,\text{dx}\,\text{dy}}{\max[\text{n}^{2}\text{(x,y)}|\text{E(x,y)}|^{2}]_\text{mode}} .
\end{equation}

\noindent  For the TE0 mode confined in the 54 nm thick, 4 $\mu$m wide $\text{Si}_{3}\text{N}_{4}$ sensing arm, the simulated values are: $$\max[\text{n}^{2}\text{(x,y)}|\text{E(x,y)}|^{2}]_\text{mode} = 4.0401,$$
$$\iint_\text{mode} \text{n}^{2}\text{(x,y)}|\text{E(x,y)}|^{2} \,\text{dx}\,\text{dy}  = 1.3192\times 10^{-12}\, \text{m}^{2},$$ and $$\iint_\text{sol} \text{n}^{2}\text{(x,y)}|\text{E(x,y)}|^{2} \,\text{dx}\,\text{dy} = 2.112\times10^{-13}\, \text{m}^{2}.$$

\noindent Molar mass ($\text{M}_{\text{m}}$) of glucose $(\text{C}_{6}\text{H}_{12}\text{O}_{6})$ is 180 g/mol, given the concentration, c in $\text{g}/\text{m}^{3}$, the number of molecules contributing to change in phase shift over the sensing window length ($\text{L}$) is $$\text{N}_\text{molecules} = \frac{\text{c}}{\text{M}_{\text{m}}} \text{N}_{\text{A}}\text{LA}_\text{eff, solution},$$
where $\text{N}_{\text{A}}$ is Avogadro's constant. 

\noindent For concentration in the order of 10 $\mu$g/ml, number of molecules contributing to measurable phase difference is $\sim 10^{7}$. Using $\delta \text{n}_{\text{sol}}/ \delta \text{c} \sim 1.56 \times 10^{-1} \,\text{RIU}/(\text{g/ml})$ for glucose solution and eqn. \ref{eq:1}, the relation between change in phase and glucose concentration can be found: $\delta \phi = \text{S}_{\text{wg}}(2 \pi  \delta \text{n}_{\text{sol}}  \text{L})/\lambda$, 
%$\delta \phi = \text{S}_{\text{wg}} (2 \pi)(\delta \text{c)(0.156 L)/}\lambda$, 
or $\delta \phi /\delta c  = \text{S}_{\text{wg}} \text{(2} \pi\text{)}\text{(0.156L)}/\lambda$. Given that the length of the sensing window is 12 mm, the wavelength of the laser is 633 nm and the calculated slope, $\delta \phi /\delta  c$ = 0.01062 $\times\,10^{6}$ rad/(g/ml) from Fig. \ref{result_plot}, we estimate the experimental value of $\text{S}_{\text{wg}}$ = 0.57.  

\noindent The phase LoD is defined as $\Delta \phi_{\text{limit}} = \text{k} \times \sigma_\phi$, where k is a constant that depends on the confidence level or detection criterion. Common values of k are: for a 68.3\% confidence level (1-sigma detection), k = 1; for a 95.4\% confidence level (2-sigma detection), k = 2; for a 99.7\% confidence level (3-sigma detection), k = 3. Since $\sigma_\phi \sim 0.03$ rad, therefore, for k=3, we have $\Delta \phi_{\text{limit}}$ = 0.09 rad, which corresponds to bulk refractive index change $\sim 1.32 \times 10^{-6} \,\text{RIU}$ and the capability of measuring 10 $\mu$g/ml glucose concentration.

\section{Conclusion}
We present design, simulations and fabrication for operating waveguide-based Young's interferometer in $\text{Si}_{3}\text{N}_{4}$ photonic platform for applications in molecules concentration measurement. We present numerical results of mode indices in different configurations of the core waveguide. We showed the optimum thickness of the strip waveguide in the fundamental TE mode operation of the sensing arm waveguide. We demonstrate bulk sensing measurements using glucose solution with different concentrations and the chip producing high-quality fringes in output. The phase estimations from each fitted interference pattern have a low error due to a minimum 0.75 visibility of fringes and less noisy interference in general. The experimental sensitivity of the sensor is found to be 0.57. The sensor's refractive index change detection limit is $\sim 1.32 \times 10^{-6}$ RIU. Currently, CMOS imaging sensors enable low-cost detection schemes combined with Young's interferometer for arbitrary phase shift detection.
\\

\noindent
This work can easily be extended to measuring the bulk concentration of antibodies and biomolecules in an aqueous solution and a thin layer of biomolecules captured on the functionalized surface of a core waveguide.

\section{Funding}
S.D. is supported by Herman F. Heep and Minnie Belle Heep Texas A\&M University Endowed Fund held/administered by the Texas A\&M Foundation. We want to thank the Robert A. Welch Foundation (grants A-1261 and A-1547), the DARPA PhENOM program, the Air Force Office of Scientific Research (Award No. FA9550-20-10366), and the National Science Foundation (Grant No. PHY-2013771). This material is also based upon work supported by the U.S. Department of Energy, Office of Science, Office of Biological and Environmental Research under Award Number DE-SC-0023103, DE-AC36-08GO28308.

\section{Acknowledgments}

Fabrication of photonic chips was performed at the Aggiefab facility of Texas A\&M University.

\section{Disclosures}

The authors declare that they have no known competing financial interests or personal relationships that could have appeared to influence the work reported in this paper.

\section{Data Availability}

Data underlying the results presented in this paper are not publicly available but can be obtained from the authors upon reasonable request.

%%%%%%%%%%%%%%%%%%%%%%% References %%%%%%%%%%%%%%%%%%%%%%%%%

%%%%%%%%%% using BibTeX:
\bibliography{sample}

\begin{thebibliography}{10}
\newcommand{\enquote}[1]{``#1''}

\bibitem{MolinaFernndez2019}
Ã.~Molina-Fern\'{a}ndez, J.~Leuermann, A.~Ortega-Mo\~{n}ux, \emph{et~al.}, \enquote{Fundamental limit of detection of photonic biosensors with coherent phase read-out,} {\protect\JournalTitle{Optics Express}} \textbf{27}, 12616 (2019).

\bibitem{Mehrotra2016}
P.~Mehrotra, \enquote{Biosensors and their applications - a review,} {\protect\JournalTitle{Journal of Oral Biology and Craniofacial Research}} \textbf{6}, 153--159 (2016).

\bibitem{Bhalla2016}
N.~Bhalla, P.~Jolly, N.~Formisano, and P.~Estrela, \enquote{Introduction to biosensors,} {\protect\JournalTitle{Essays in Biochemistry}} \textbf{60}, 1--8 (2016).

\bibitem{Haleem2021}
A.~Haleem, M.~Javaid, R.~P. Singh, \emph{et~al.}, \enquote{Biosensors applications in medical field: A brief review,} {\protect\JournalTitle{Sensors International}} \textbf{2}, 100100 (2021).

\bibitem{Alemdar2024}
S.~Alemdar, N.~Pekel~Bayramgil, and S.~Karakus, \emph{Applications of Cutting-Edge Biosensors in Healthcare and Biomedical Research} (IntechOpen, 2024).

\bibitem{Brandenburg1994}
A.~Brandenburg and R.~Henninger, \enquote{Integrated optical young interferometer,} {\protect\JournalTitle{Applied Optics}} \textbf{33}, 5941 (1994).

\bibitem{Liu1992}
Y.~Liu, P.~Hering, and M.~O. Scully, \enquote{An integrated optical sensor for measuring glucose concentration,} {\protect\JournalTitle{Applied Physics B Photophysics and Laser Chemistry}} \textbf{54}, 18--23 (1992).

\bibitem{Ymeti2006}
A.~Ymeti, J.~Greve, P.~V. Lambeck, \emph{et~al.}, \enquote{Fast, ultrasensitive virus detection using a young interferometer sensor,} {\protect\JournalTitle{Nano Letters}} \textbf{7}, 394--397 (2006).

\bibitem{Estevez2011}
M.~Estevez, M.~Alvarez, and L.~Lechuga, \enquote{Integrated optical devices for lab-on-a-chip biosensing applications,} {\protect\JournalTitle{Laser \& Photonics Reviews}} \textbf{6}, 463--487 (2011).

\bibitem{Tan2015}
C.-Y. Tan and Y.-X. Huang, \enquote{Dependence of refractive index on concentration and temperature in electrolyte solution, polar solution, nonpolar solution, and protein solution,} {\protect\JournalTitle{Journal of Chemical \& Engineering Data}} \textbf{60}, 2827--2833 (2015).

\bibitem{Leuermann2019}
J.~Leuermann, A.~Fern\'{a}ndez-Gavela, A.~Torres-Cubillo, \emph{et~al.}, \enquote{Optimizing the limit of detection of waveguide-based interferometric biosensor devices,} {\protect\JournalTitle{Sensors}} \textbf{19}, 3671 (2019).

\bibitem{Butt2023}
M.~A. Butt, N.~L. Kazanskiy, S.~N. Khonina, \emph{et~al.}, \enquote{A review on photonic sensing technologies: Status and outlook,} {\protect\JournalTitle{Biosensors}} \textbf{13}, 568 (2023).

\bibitem{Sinatkas2021}
G.~Sinatkas, T.~Christopoulos, O.~Tsilipakos, and E.~E. Kriezis, \enquote{Electro-optic modulation in integrated photonics,} {\protect\JournalTitle{Journal of Applied Physics}} \textbf{130} (2021).

\bibitem{Chen2023}
X.~Chen, J.~Lin, and K.~Wang, \enquote{A review of silicon-based integrated optical switches,} {\protect\JournalTitle{Laser \& Photonics Reviews}} \textbf{17} (2023).

\bibitem{Altug2022}
H.~Altug, S.-H. Oh, S.~A. Maier, and J.~Homola, \enquote{Advances and applications of nanophotonic biosensors,} {\protect\JournalTitle{Nature Nanotechnology}} \textbf{17}, 5--16 (2022).

\bibitem{Miyazaki2017}
C.~M. Miyazaki, F.~M. Shimizu, and M.~Ferreira, \emph{Surface Plasmon Resonance (SPR) for Sensors and Biosensors} (Elsevier, 2017), pp. 183--200.

\bibitem{Taniguchi2016}
T.~Taniguchi, A.~Hirowatari, T.~Ikeda, \emph{et~al.}, \enquote{Detection of antibody-antigen reaction by silicon nitride slot-\-ring biosensors using protein g,} {\protect\JournalTitle{Optics Communications}} \textbf{365}, 16--23 (2016).

\bibitem{Inan2017}
H.~Inan, M.~Poyraz, F.~Inci, \emph{et~al.}, \enquote{Photonic crystals: emerging biosensors and their promise for point-of-care applications,} {\protect\JournalTitle{Chemical Society Reviews}} \textbf{46}, 366--388 (2017).

\bibitem{Gowdhami2022}
D.~Gowdhami, V.~R. Balaji, M.~Murugan, \emph{et~al.}, \enquote{Photonic crystal based biosensors: an overview,} {\protect\JournalTitle{ISSS Journal of Micro and Smart Systems}} \textbf{11}, 147--167 (2022).

\bibitem{Singh2021}
R.~Singh, Y.~Nie, M.~Gao, \emph{et~al.}, \enquote{Inverse design of photonic meta-structure for beam collimation in on-chip sensing,} {\protect\JournalTitle{Scientific Reports}} \textbf{11} (2021).

\bibitem{Chung2021}
H.~Chung and S.~V. Boriskina, \enquote{Inverse design of a single-frequency diffractive biosensor based on the reporter cleavage detection mechanism,} {\protect\JournalTitle{Optics Express}} \textbf{29}, 10780 (2021).

\bibitem{DidariBader2024}
A.~Didari-Bader, S.~Pelton, and N.~Mohammadi~Estakhri, \enquote{Inverse-designed integrated biosensors,} {\protect\JournalTitle{Optical Materials Express}} \textbf{14}, 1710 (2024).

\bibitem{Chung2022}
H.~Chung, J.~Park, and S.~V. Boriskina, \enquote{Inverse-designed waveguide-based biosensor for high-sensitivity, single-frequency detection of biomolecules,} {\protect\JournalTitle{Nanophotonics}} \textbf{11}, 1427--1442 (2022).

\bibitem{Martens2019}
D.~Martens and P.~Bienstman, \enquote{Study on the limit of detection in mzi-based biosensor systems,} {\protect\JournalTitle{Scientific Reports}} \textbf{9} (2019).

\bibitem{Luo2022}
Y.~Luo, R.~Yin, L.~Lu, \emph{et~al.}, \enquote{Asymmetric mach-zehnder interferometer-based optical sensor with characteristics of both wavelength and temperature independence,} {\protect\JournalTitle{Journal of Optics}} \textbf{52}, 1008--1021 (2022).

\bibitem{GonzlezGuerrero2016}
A.~B. Gonz\'{a}lez-Guerrero, J.~Maldonado, S.~Herranz, and L.~M. Lechuga, \enquote{Trends in photonic lab-on-chip interferometric biosensors for point-of-care diagnostics,} {\protect\JournalTitle{Analytical Methods}} \textbf{8}, 8380--8394 (2016).

\bibitem{Alajlan2020}
A.~Alajlan, M.~Khurana, X.~Liu, \emph{et~al.}, \enquote{Free-standing silicon nitride nanobeams with an efficient fiber-chip interface for cavity qed,} {\protect\JournalTitle{Optical Materials Express}} \textbf{10}, 3192 (2020).

\bibitem{Xiang2022}
C.~Xiang, W.~Jin, and J.~E. Bowers, \enquote{Silicon nitride passive and active photonic integrated circuits: trends and prospects,} {\protect\JournalTitle{Photonics Research}} \textbf{10}, A82 (2022).

\bibitem{Wang2012}
M.~Wang, J.~Hiltunen, C.~Liedert, \emph{et~al.}, \enquote{An integrated young interferometer based on uv-imprinted polymer waveguides for label-free biosensing applications,} {\protect\JournalTitle{Journal of the European Optical Society-Rapid Publications}} \textbf{7}, 12019 (2012).

\bibitem{Zhou2018}
C.~Zhou, M.~K. Hedayati, and A.~Kristensen, \enquote{Multifunctional waveguide interferometer sensor: simultaneous detection of refraction and absorption with size-exclusion function,} {\protect\JournalTitle{Optics Express}} \textbf{26}, 24372 (2018).

\bibitem{Wong2019}
W.~R. Wong and P.~Berini, \enquote{Integrated multichannel young's interferometer sensor based on long-range surface plasmon waveguides,} {\protect\JournalTitle{Optics Express}} \textbf{27}, 25470 (2019).

\bibitem{Watchi2018}
J.~Watchi, S.~Cooper, B.~Ding, \emph{et~al.}, \enquote{Contributed review: A review of compact interferometers,} {\protect\JournalTitle{Review of Scientific Instruments}} \textbf{89} (2018).

\bibitem{Milvich2018}
J.~Milvich, D.~Kohler, W.~Freude, and C.~Koos, \enquote{Surface sensing with integrated optical waveguides: a design guideline,} {\protect\JournalTitle{Optics Express}} \textbf{26}, 19885 (2018).

\bibitem{Agrawal2019}
G.~P. Agrawal, \emph{Pulse propagation in fibers} (Elsevier, 2019), pp. 27--55.

\end{thebibliography}

\end{document}